\documentclass[twocolumn]{article}
\usepackage[utf8]{inputenc}
\usepackage{amsmath}
\usepackage{graphicx}
\graphicspath{ {./images/} }

\usepackage{algorithm}
\usepackage[noend]{algpseudocode}

\usepackage{listings}
\lstdefinestyle{mystyle}{
 basicstyle=\ttfamily\small
}
\usepackage{multicol}
\usepackage{verbatim}
\usepackage{appendix}
\usepackage{makecell}

\title{Toward Lattice QCD on \\ Billion Core Approximate Computers}

\author{\\
Alexandra Bates 
\thanks{We thank Andrew Pochinsky of the MIT Center for Theoretical Physics for his guidance and encouragement. All views and errors are those of the authors. Alexandra Bates had primary responsibility for the software aspects of this work. Joseph Bates had primary responsibility for hardware aspects.} 
\\
Commonwealth School, Boston, MA \\
\emph{sbates10@icloud.com}
\and \\ 
Joseph Bates \\
Singular Computing LLC, Cambridge, MA}

\date{October 29, 2020 \vspace{1em}}

\begin{document}

\maketitle

\begin{abstract}
We present evidence of the feasibility of using billion core approximate computers to run simple U(1)$\sigma$ models, and discuss how the approach might be extended to Lattice Quantum Chromodynamics (LQCD) models. This work is motivated by the extreme time, power, and cost needed to run LQCD on current computing hardware. We show that, using massively parallel approximate hardware, at least some models can run with great speed and power efficiency without sacrificing accuracy. As a test of accuracy, a $32 \times 32 \times 32$ U(1)$\sigma$ model yielded similar results using floating point and approximate representations for the spins. A 20 million point 3D model, run on a 34,000-core single-board prototype approximate computer, showed encouraging accuracy with a $\sim\!750\times$ improvement in speed and  $\sim\!2500\times$ improvement in speed/watt compared to a traditional CPU.  These results suggest there is value in future research to determine whether similar speed-ups and accuracies are possible running full LQCD on the compact billion-core approximate computing systems that are now practical.
\end{abstract}

\section{Introduction}
Lattice Quantum Chromodynamics (LQCD) is a useful model of the strong force. However, to run LQCD models requires immense computing power on a scale difficult and expensive to deliver using traditional hardware. LQCD is typically run on lattices containing between $10^8$ and $10^{12}$ points, and a typical model requires evaluating O(1000) lattice configurations and performing O(1000) measurements per configuration \cite{cite:mitqcd}.  This therefore requires up to O($10^{18}$) computational steps, where each step itself may be complex and require a large number of arithmetic operations. The sheer amount of computation necessary means that these models run slowly while consuming large amounts of energy.  

This time and energy cost severely limits the size of possible models and the number of configurations which can be explored. This in turn limits physicists' knowledge of quantum chromodynamics. If the computational constraint could be lessened, we would be able to run significantly more sophisticated LQCD models, allowing for further progress in the field.

The time and energy costs can be lessened by using nontraditional hardware. Specifically, herein we consider the feasibility of running U(1)$\sigma$ models using low precision high dynamic range (``approximate'') hardware: in particular, computers developed by Singular Computing of Cambridge, MA. These computers are fairly general purpose programmable machines that can be adapted to many types of problems. Their low precision arithmetic operations allow their arithmetic circuitry to use far fewer transistors than usual, allowing the computer as a whole to be 50 times more efficient than GPUs (graphics processing units), which currently are the most efficient general purpose machines in common use. With this approach, it is practical to build compact billion-core computers, as discussed in section \ref{bcc}.

For U(1)$\sigma$ Models, the large number of cores permits a natural mapping of the logical lattice onto the physical grid of cores. Each core contains a set of lattice points and all cores update their point sets in parallel. 

If this approach could be implemented on low precision hardware without compromising the accuracy of the final results, it would have substantial consequences for the efficiency of LQCD models and would allow larger models to be run over more iterations. In this paper, we discuss the feasibility of this approach. We explain the capabilities and limitations of Singular-style hardware, give examples of fast and accurate implementations of U(1)$\sigma$ models on approximate hardware, and present ideas for scaling up to full LQCD models. 

\section{Billion Core Computers}  
\label{bcc}

Traditional CPUs and GPUs perform high precision arithmetic, such as 32 bit and 64 bit floating point arithmetic.  Circuits for performing such arithmetic typically use hundreds of thousands of transistors, and the processing elements (``cores’’) containing such circuits typically use millions to billions of transistors.  These large cores perform a handful of arithmetic operations for each clock cycle of computer operation.

If one were to represent values as electrical currents then wire junctions could be used to sum the currents and thus perform addition.  Individual transistors operating in the sub-threshold regime could be used to compute exponentials and logarithms.  Whether this is precisely the approach one would use or not, these observations suggest that today’s silicon fabrication factories have the ability to produce chips able to perform many millions or billions of operations each clock cycle.

These ideas have been discussed for decades, for instance by Carver Mead while at Caltech \cite{cite:mead}.  They point toward a path that could yield a huge jump in computing power, which history tells us almost certainly would be enormously consequential. 

However, exploiting sub-threshold analog computing presents difficulties, especially when trying to build programmable general purpose machines.  One concern is that the arithmetic operations would be approximate - introducing possibly unacceptable error in each operation.  In addition, transistor scaling in advanced silicon fabrication processes seems to make analog computing increasingly difficult.  Thus, Mead's suggestions may or may not be fully feasible, but as a step in that direction, it demonstrably is possible to replace current digital floating point circuits with much smaller circuits that do high dynamic range approximate arithmetic.  For instance, a circuit able to perform addition, subtraction, multiplication, division, and square root can be $\sim\!\!100\times$ smaller than a typical floating point arithmetic unit.

Such arithmetic circuits produce results that are imprecise compared to traditional floating point arithmetic. For instance, although deterministic and thus reproducible, relative error of the results may frequently reach as high as 1\% compared to correct results.  Such errors may seem to make it impossible to create useful software, but that is not the case, as we shall discuss, and by accepting this trade-off it is practical to build compact programmable digital computers with millions of cores per board and hundreds of millions of cores per server rack.  Thus, compact billion core systems become practical.

With support from the U.S. Defense Advanced Research Projects Agency (DARPA), Singular Computing has taken steps down this path.  It has built and operates first generation prototypes, such as single board computers with  $\sim\!34,000$ cores, which we shall describe.

Once the arithmetic unit becomes extremely small, it is necessary to reduce the size of the other components of the computing system to avoid the consequences of Amdahl's Law, which points out that the most inefficient components of a system dominate its overall cost.  Singular's S1 architecture simplifies a typical modern computing architecture by returning to the 2D mesh organization from decades ago, such as was used in the MPP, DAP, part of the Connection Machine, and other machines \cite{cite:cypher}.

\begin{figure}
\begin{center}
\includegraphics[scale=0.4]{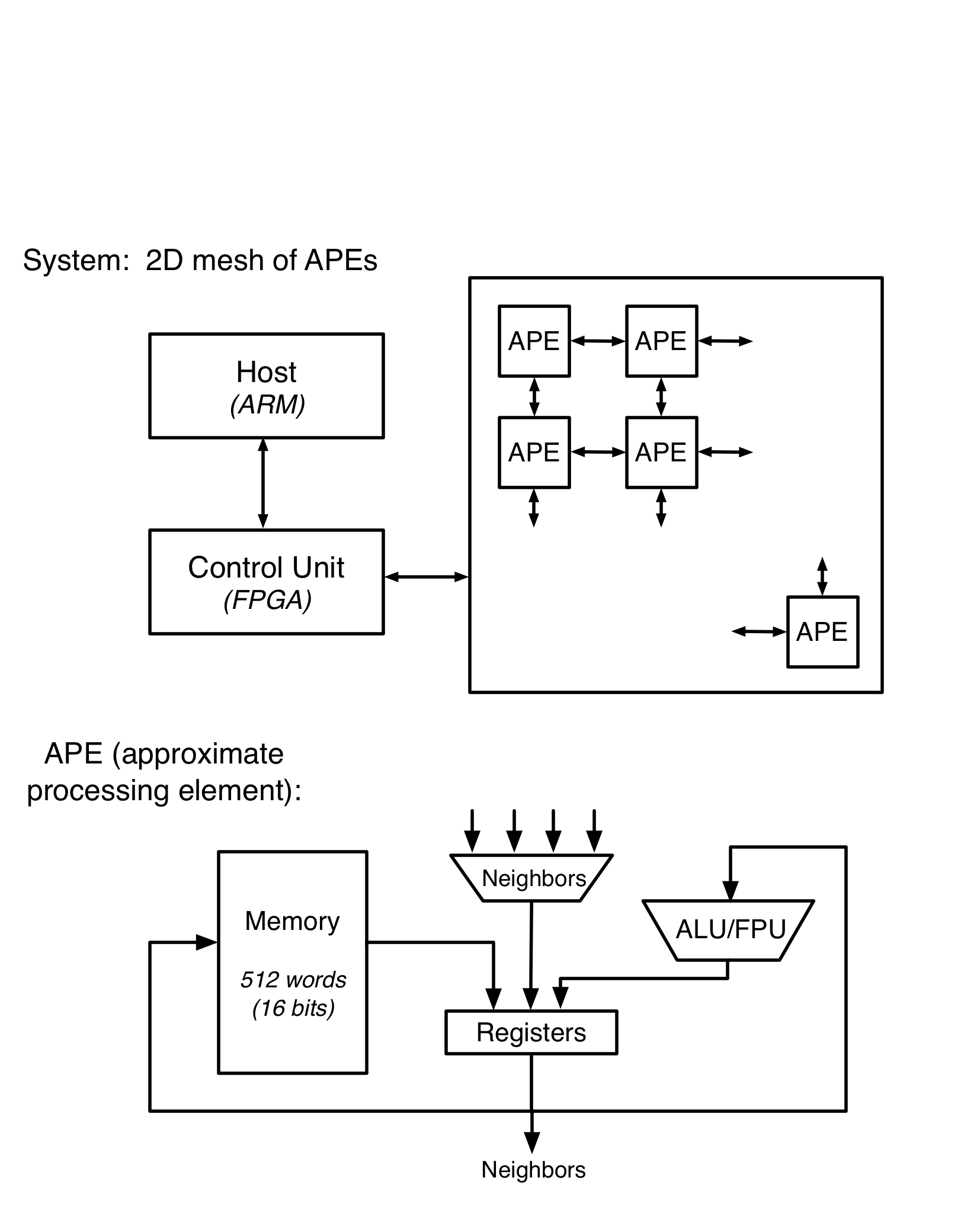}
\end{center}
\caption{Architecture of S1 system, including mesh and individual processing elements.}
\label{fig:S1architecture}
\end{figure}

The overall structure of the S1 prototype is shown in the upper part of Figure \ref{fig:S1architecture}.  There is a conventional host CPU, connected to a Control Unit, which itself is a simple traditional computer.  The bulk of the user's code runs on the host in the usual manner.  But part of that code's job is to create other code (called a ``kernel'') that runs on the Control Unit.  The Control Unit in turn broadcasts instructions contained in the kernel to an attached array of cores, called approximate processing elements and denoted APEs in the figure. This arrangement, in which a host CPU collaborates and communicates with a second processor, is similar to how graphics processing units (GPUs) are connected to host CPUs.

Each APE of the S1 contains 512 16-bit words of memory, 8 16-bit registers, circuits to perform arithmetic and logical operations, connections to its four nearest neighbors (to the North, East, South, and West - called NEWS connections), and global connections back toward the Control Unit (not shown in the figure).  The supported logical operations include $and, or, not$ along with variable length shifts.  Exact 16-bit integer addition and subtraction is supported.  Approximate  (1\% relative error) addition, subtraction, multiplication, division, and square root is supported on high dynamic range ($10^{-18} \cdots 10^{18}$) values represented as 16-bit fixed-point logarithms.

In typical use, the mesh of APEs all receive, at any given moment, the same instruction from the Control Unit.  They all perform that instruction at the same time.  However, the instruction may refer to local information, such as the content of registers, and that information can be used to specify a memory address, or one of the APE's neighbors, or even another instruction, so it is in fact oversimplified to say every APE performs the same operation at each moment.  However, in a broad sense this design is referred to as a Single Instruction Stream Multiple Data Stream (SIMD) array processor.

Silicon chips are very nearly 2D structures.  They can have perhaps a dozen layers of metal wiring crisscrossing above the transistors embedded in the silicon.  Yet chips can extend for hundreds of thousands of transistors in the X and Y directions in the plane of the chip.  Thus, their aspect ratio is something like 10,000 to 1 and they are effectively 2D entities.  This means that if we ask hardware designers to build sophisticated communication architectures that are significantly different from a 2D mesh, they will do it, but the sophisticated architecture will in fact be embedded in a 2D material, and much silicon and wiring will be spent making it appear to the user that silicon is not essentially 2D.

Singular feels that a better approach, for users who really want to make full use of what our current silicon technology can provide, is to expose to software the reality that hardware is 2D and allow (and indeed force) programmers to build specialized communication methods (perhaps embedded in software libraries) that are optimal for their individual problems. 

Making this choice, along with the choice to do approximate arithmetic, means that individual silicon chips can hold hundreds of thousands of programmable cores, allowing at least some problems to be implemented in a way that comes closer to Mead's vision of each transistor and wire junction performing useful work.

The S1 prototype chip was built using a 40nm silicon fabrication process and a $25mm^2$ silicon die.  It contains 2112 APEs, in a $48 \times 44$ grid.  As of 2020, new mobile phones use 5nm silicon.  GPUs use dies of roughly $600mm^2$.  Thus, using the most modern process of the year 2020, and a large die, one could increase the APE count by a factor approaching $(40/5)^2 \times (600/25) = 1536$, and therefore place at least hundreds of thousands of cores on a chip and millions of cores on a board containing a mesh of chips.  The feasibility of scaling algorithms to such sized machines depends on the algorithms, but the extreme data parallelism in certain aspects of LQCD make it an appealing candidate.

The 34K-core S1 prototype system is composed of an ARM CPU running Linux, a Control Unit (programmed into an FPGA), and a 4x4 array of S1 chips.  The Control Unit can communicate with each chip individually, or can broadcast instructions and data to all the chips in parallel.  The chips contain internal communication paths, discussed below.  These paths extend out to the edges of the chips, where they can connect to corresponding paths in the neighboring chips, so that the entire array of $(4 \times 48) \times (4 \times 44) = 33,792$ APEs can operate as a single uniform mesh. 

Programming the Control Unit, i.e. building the kernels, is easier if there is a high level language available.  Singular developed a simple such language, called \emph{Nova}, for this purpose.  Nova code is embedded in the host machine's C code, and a runtime compiler is available to translate the constructed Nova code into machine code for the Control Unit.  This approach is similar to the OpenCL language for GPUs and to the TensorFlow language for TPUs (tensor processing units for deep neural nets).

The details of this process, and the syntax and semantics of Nova, are not essential aspects of this paper.  Appendix \ref{appendix:matmul}, discussed below, shows an example of how C and Nova are used to program the accelerator and have the CPU host and the Control Unit communicate.

There are several data communication paths across the S1 mesh.  Local communication between neighbors is accomplished using \emph{Get} instructions.  When the Control Unit broadcasts a \emph{Get} instruction, it specifies a NEWS (North, East, West, South) direction from which to get a single value. Each core looks at (a register inside) its neighbor in that direction, grabs the value, and brings it to (a register inside) itself.  The direction can be specified as a constant, or the core can be instructed to find the direction in a local variable (e.g. in a register).  This allows complex patterns of data motion to be programmed, such as local or global cycles, serpentine ``snakes'', etc.

The S1 supports certain kinds of global communication.  The Control Unit is able to broadcast instructions and data to any subset of the APEs, specifying those APEs using a mix of chip and APE row and column numbers along with locally computed flags inside each APE that mark it as listening or ignoring the broadcasts.  Data within the APEs can also be read back across the rows of each chip, then upward along the leftmost edge of the chip, and finally back to the Control Unit.  The value read back can be from a particular APE, again specified using rows, columns, and local flags.  In addition, if multiple APEs are specified, their respective values (16-bit words) can be bitwise OR'ed into a single value read back to the Control Unit.

The global OR reduction is fast, typically taking a few cycles to complete.  More complex reductions are done by interleaving the Get instruction with the instruction(s) needed for the desired reduction.  For instance, to do a global sum reduction, every APE will Get a partial sum from its right, add its own value to the sum, and repeat this 48 times (the width of the mesh on each chip).  The leftmost APEs will then contain the partial sums of their entire rows, and we can repeat the process vertically 44 times so that the upper left corner APE will have the sum of the whole chip.  At this point the Control Unit can read back and compute the sum of these values from all 16 chips and, for instance, broadcast the global sum (or some function of the global sum) back to all 34K APEs.

The global sum reduction thus takes a few hundred cycles on the S1 prototype.  The S2 production system being developed speeds up this operation by an order of magnitude, and supports more flexible semi-local reductions and broadcasts.

To make the ideas above more concrete, let us consider an example of matrix multiplication that uses the Singular architecture, as shown in Algorithm \ref{alg:matmul}.  At each core in the mesh we declare a 3x3 matrix of approximate values.  The algorithm calculates, in parallel at each core located at $(i,j)$, the product of the matrix at core $(i-1,j)$ with the matrix at core $(i,j)$, storing the result in core $(i,j)$.

This example shows the SIMD programming model and local communication.  While the code appears to be a sequential single-core algorithm, it is in fact being broadcast to all 34K cores in the machine, and they all perform their own matrix multiplications in parallel.  The only unusual operation is Get, which brings to each core a single value from its neighbor to the west.  Thus, after O($3^3$), i.e. several dozen, operations, all 34,000 3x3 matrices have been multiplied by their neighbor matrices to the west.

Appendix \ref{appendix:matmul} presents code, written in C with embedded Nova code for the accelerator mesh, that implements this algorithm on hardware, but where the matrices contain complex numbers rather than real numbers.

For this example, we assumed the mesh is connected as a torus, with the westmost cores connected to the eastmost cores (and topmost to bottommost).  The code in Appendix \ref{appendix:matmul} declares a machine (using the \emph{scInitializeMachine} function call) organized in this way, and this is supported while emulating the S1 on a CPU.  The actual 34K-core prototype hardware does not contain these torus connections, but the 20M point model in section \ref{performance} shows a way to obtain torus behavior on hardware without torus connections.

\begin{algorithm}
\begin{algorithmic} 
\vspace{1em}
\State Declare Approximate A[3,3], B[3,3], C[3,3]
\\
\State // Initialize A and B
\ForAll{$i,j \in 1..3$}
  \State $A_{ij} \leftarrow$ random approximate value
  \State $B_{ij} \leftarrow$ random approximate value
\EndFor
\State
\State // Get the A array from the neighbor to the left
\ForAll{$i,j \in 1..3$}
  \State $a \leftarrow A_{ij}$
  \State Get(a, {West})
  \State $A_{ij} \leftarrow a$
\EndFor
\State
\State // Multiply A by B to get C
\ForAll{$i,j \in 1..3$}
  \State $c_{ij} \leftarrow 0$
  \ForAll{$k \in 1..3$}
    \State $C_{ij} \leftarrow approxAdd( C_{ij}, approxMul(A_{ik}, B_{kj}))$
  \EndFor
\EndFor
\State
\caption{Multiplication of 3x3 approximate-valued matrices, stored in adjacent cores, on S1 prototype, in parallel on all 34,000 cores}
\label{alg:matmul}
\end{algorithmic}
\end{algorithm}

\section{Accuracy of a 3D U(1)$\sigma$ Model using Approximate Arithmetic}
\label{accuracy}

For the hardware described above to be useful, the results of models run on it must be sufficiently accurate. To judge the accuracy of approximate arithmetic for our application, we developed two implementations of a U(1)$\sigma$ model, storing important information as 32-bit floating point values in one and as 16-bit approximate values in the other. 

The U(1)$\sigma$ model is a limiting case of the Heisenberg spin model. Our particular model contains a three-dimensional lattice of spins.

The model is intended to reflect the physics of a lattice, which typically will exhibit phase transitions at certain temperatures. Accurately finding the temperatures at which transitions occur is the main purpose of the model. This is done by varying the coupling constant $\beta = 1/temperature$, and observing the energy of the lattice. The transition is considered to occur at the point at which the energy has the greatest derivative. 

This algorithm requires representation of spins. A spin is an $(x, y)$ pair of Cartesian coordinates representing a 2D vector on the unit circle. We represented the numbers in two ways: as 32 bit floating point values or 16 bit approximate values. We now discuss a few key features of the floating point and approximate implementations.  Appendix \ref{appendix:floatApproxCode} provides more detail, showing the code that implements the model.

First, a few remarks on the general algorithm.  Modeling the lattice behavior involves repeatedly considering possible changes to the lattice point spins, and updating those spins in a probabilistic way based on the energy change resulting from replacing current spins with the possible new spins.  This is a Markov Chain Monte Carlo algorithm, as described for example in  \cite{cite:xy}

The idealized algorithm updates all spins in parallel. However, this poses an implementation problem if we want to avoid increasing memory demands in what is already a memory hungry task. Since each spin's updated value depends on the pre-update values of its neighbors, updating every spin simultaneously would require storing the old and new values for each spin, thereby doubling the necessary storage. An effective alternative is to update the spins in a checkerboard pattern. On the even steps, each red spin is updated, while on the odd steps, each black spin is updated. 

To update a spin, a random new spin is computed and the consequent change in energy $\Delta E$ is calculated in the following manner. Each point in a 3D lattice has six neighbors, two in each dimension. The energy of a lattice of spins is the sum of the energies of the edges connecting each of its points.  This is half the energy of each of its points (because each link will be counted twice, once for each end of the link), where the energy of a point at $(i, j, k)$ is defined by 
\begin{equation} \label{energy}
    E = -J \times \sum_{n \in neighbors(i, j, k)} s_n \cdot s_{ijk}
\end{equation}
and where each $s$ is a spin, $s_n \cdot s_{ijk}$ is the dot product of the 2-vectors, and $J$ is a constant. The new spin is accepted with probability
\begin{equation} \label{exp}
    p = 1/z \times e ^ {- \beta \Delta E / K} 
\end{equation}
where $z$ is a normalization factor and K is Boltzmann's constant. 

Given these equations, the floating point algorithm is straightforward. The only subtlety is that in order to update the model at a lattice point, we must compare the energy corresponding to the current spin with the energy corresponding to some randomly chosen new spin. Therefore, we must find a way to uniformly sample the distribution of possible angles on the unit circle. 

To do so, we choose a random pair $(x,y)$ in a square of edge length 2 centered at the origin and use this pair to determine a random point inside the unit circle. We cannot simply scale the vector defined by this point onto the unit circle, because that would yield non-uniform sampling of the distribution. Instead, we scale it if it falls on or within the unit circle, and discard it if it does not. In our implementation, we repeat this up to five times. If by the fifth time, none of the randomly generated pairs have fallen inside or on the unit circle, we simply scale the fifth pair, ignoring the non-uniform sampling. This case arises with probability $(1 - \pi / 4)^5 = .0005$, which we take as sufficiently rare as to cause minimal skew in our feasibility results. (And of course the five iterations could be increased, or another sampling method used.)

For every lattice point, we now have a new possible spin. We compute the change in energy resulting from changing the current spin to the new one. If the change in energy is negative, we accept the new spin. If it is positive, we accept with probability given by equation \ref{exp}. 

Code describing in detail the process for updating spins, and for sweeping $\beta$ as discussed below to determine the temperature of any phase transition, is included in Appendix \ref{appendix:floatApproxCode}

We now wish to adapt this algorithm to run using approximate XY values for the spins, instead of floating point values.  Singular has found that some algorithms require significant modification to work well using approximate arithmetic. However, as the results presented below suggest, the straightforward algorithm produces good results if we simply store each spin as a pair of approximate values and throughout the algorithm use approximate arithmetic operations instead of floating point arithmetic operations.

In order to evaluate the quality of results, however, there is an adjustment we make. Singular's approximate values are accurate to about $1\%$. This means a spin that is supposed to lie on the unit circle might be noticeably incorrect, such as having a length of $0.97$ instead of $1$. The approximate variant of the Monte Carlo algorithm proceeds through it's entire series of spin update steps using approximate values.  Then, in order to report the actual energy of the lattice at each step, we perform a post-processing step that converts each spin from an approximate to a floating point pair, scales the pair to lie on the unit circle using floating point arithmetic, and then calculates the energy at each point per equation \ref{energy} using floating point arithmetic.

We now discuss the higher level algorithm that allows us to understand the phase change behavior of the lattice.  Our goal is to obtain a graph of the energy per lattice edge versus $\beta$.

The algorithm consists of considering various values of $\beta$, and for each value performing updates until the energy settles into a stable range. We choose in advance the number of updates which we believe will be sufficient to reach this equilibrium state. In order to determine whether our choice is sufficiently large to produce accurate measurements, we run $\beta$ up from $\beta_{min}$ to $\beta_{max}$ and then down from $\beta_{max}$ to $\beta_{min}$. For each $\beta$ we perform updates, our hypothesized sufficient number of times, and compare the equilibrium energies for each $\beta$ from the up and down runs. If the energies are nearly identical for each $\beta$, then our choice is large enough.

\begin{figure}
\begin{center}
\includegraphics[scale=0.33]{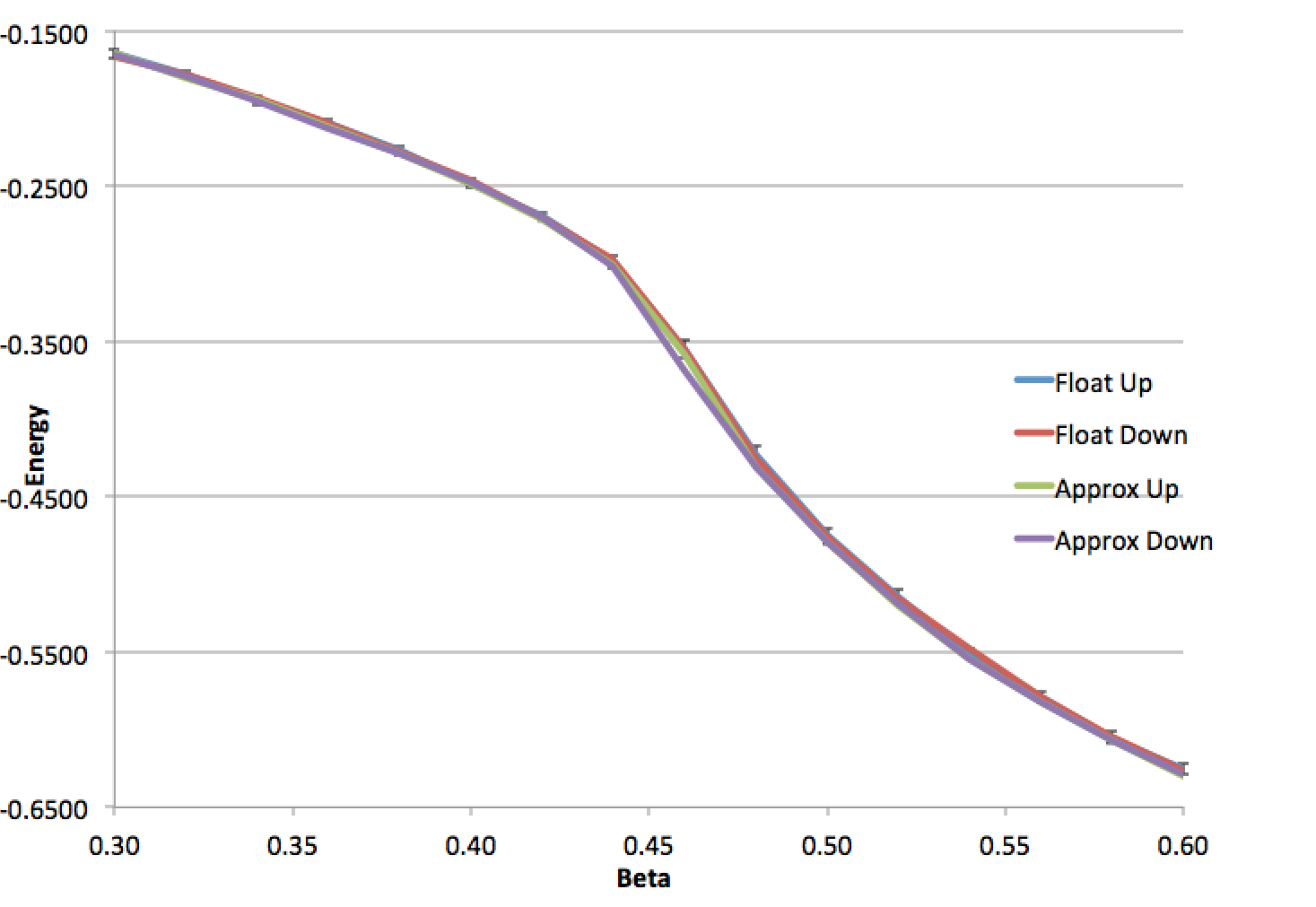}
\end{center}
\caption{Average equilibrium energy per lattice edge (y-axis) vs.~coupling constant $\beta$ (x-axis), for floating point and approximate representations of lattice values and over increasing and decreasing coupling constants.  The error bars show one standard deviation above and below the means.}
\label{fig:energyFloatApprox}
\end{figure}

\begin{figure} 
\begin{center}
\includegraphics[scale=0.33]{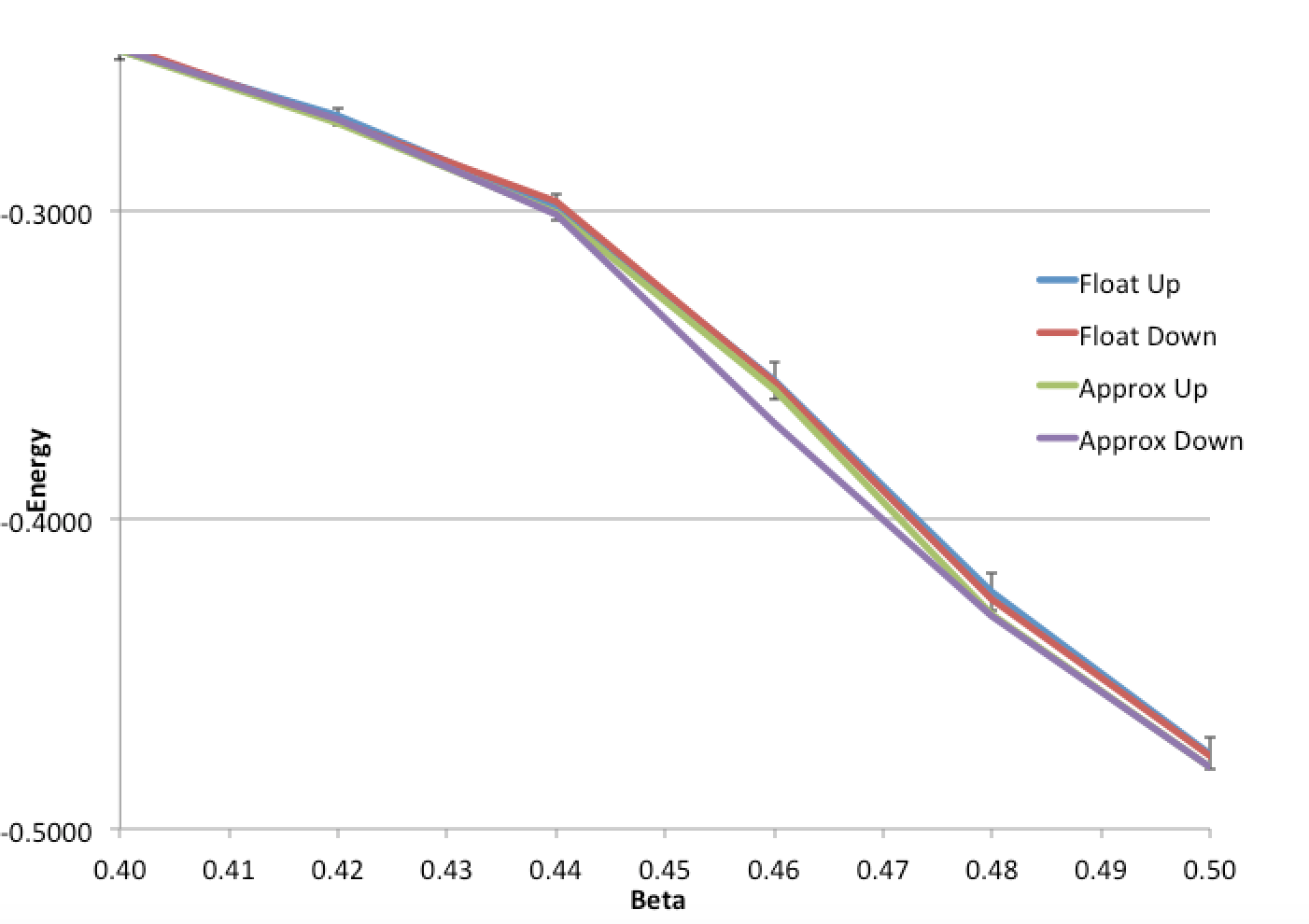}
\end{center}
\caption{The same curves as in Figure \ref{fig:energyFloatApprox}, but with the region of the phase transition enlarged.}
\label{fig:zoomedEnergyFloatApprox}
\end{figure}

In the results presented here, we ran the algorithm on a $32 \times 32 \times 32$ point lattice, varying $\beta$ from 0.3 to 0.6 by 0.02. At each $\beta$, we updated the lattice 1,000 times to attempt to reach equilibrium, and then 2,000 more times for measurements, recording the total energy after each update, to collect a representative history of energies. These energies are processed as explained below to find the average equilibrium energy per lattice edge, as shown in the graphs in Figures \ref{fig:energyFloatApprox} and \ref{fig:zoomedEnergyFloatApprox}.

When we run the measurement phase, performing 2000 spin updates and calculating their respective configuration energies, the energies of configurations that are close in time will tend to be close in value. That is, the configurations and energies are correlated.  This makes a simple calculation of standard deviation misleading.

\begin{figure}
\begin{center}
\includegraphics[scale=0.33]{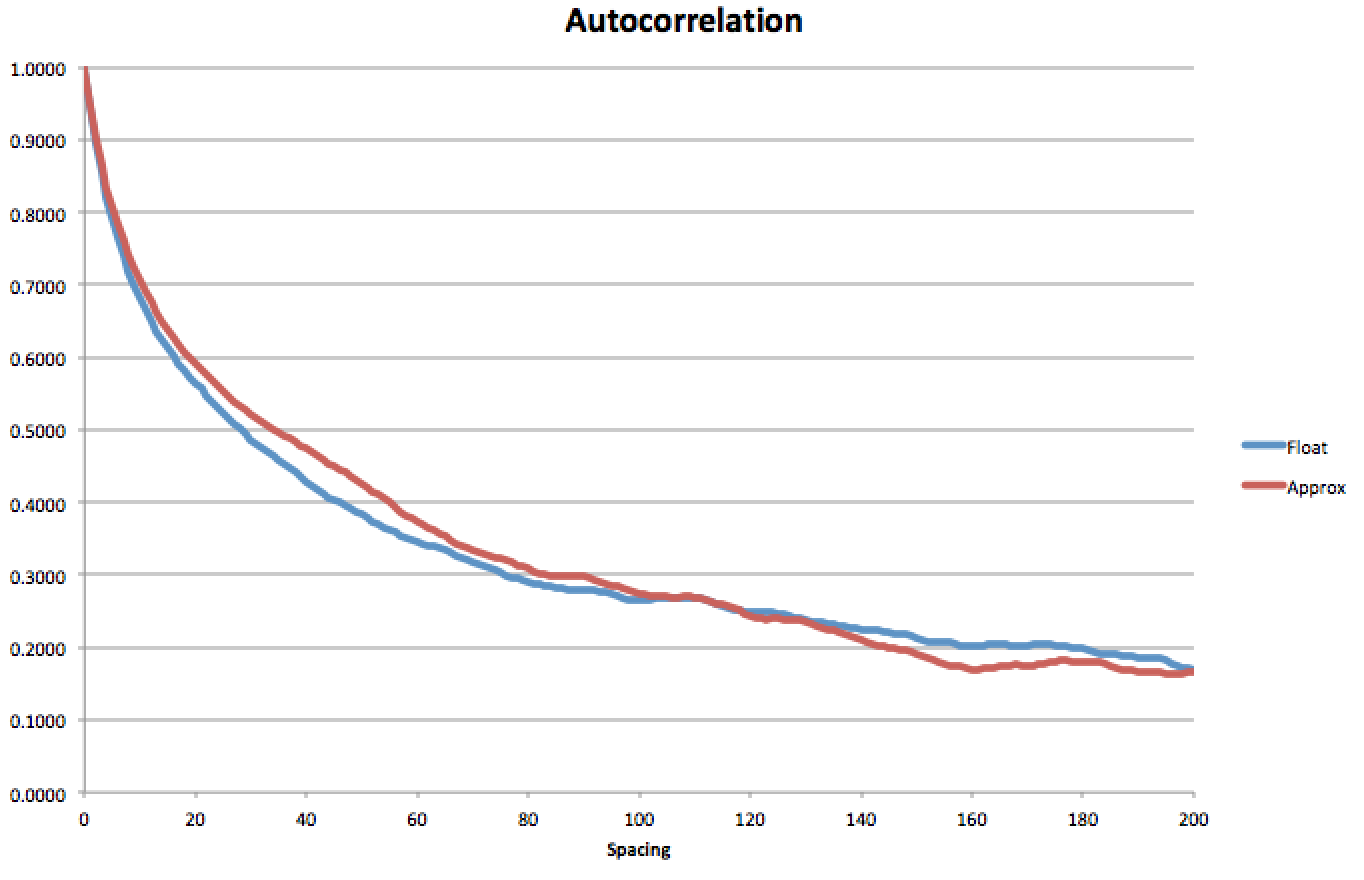}
\end{center}
\caption{Autocorrelation of energies, for floating point and approximate representations of lattice points, for spacings of 1 to 200.} 
\label{fig:autocorrelation}
\end{figure}

To avoid this problem, We plotted and analyzed the auto-correlation of the energies, as shown in Figure \ref{fig:autocorrelation}.  The phase transition occurs near $\beta = 0.46$, so we chose that $\beta$ value to do the analysis.  We ran 20,000 measurement steps.  For each spacing ${j}$ from 1 to 200, we computed the auto-correlation of the energies for measurement $i$ and measurement $i+j$.  For a spacing of about 100 steps, the energy values de-correlated. We then averaged one out of every 100 energy measurements for each $\beta$ to find the average equilibrium energy of the lattice at that $\beta$.  Dividing the lattice energy by the number of links ($3 \times 32^3 = 98,304$) gave the average equilibrium energy per lattice link.  We also computed corresponding standard deviations.

As shown in Figure \ref{fig:energyFloatApprox}, which shows average energies with error bars showing corresponding standard deviations, the differences in average equilibrium energy per lattice link between the sweeps over increasing and decreasing beta, using floating point spins, are within one standard deviation.  This provides confidence in the accuracy of our algorithm. Additionally, the differences in average equilibrium energy per lattice link between the floating point and approximate representations of lattice values are within one standard deviation on both the increasing and decreasing beta sweeps. This shows that, at least in the model considered here, using approximate values and arithmetic can give similarly accurate results as using floating point values and arithmetic. Since the U(1)$\sigma$ Model uses some of the same operations and patterns as common LQCD models, these results suggest that approximate values may be promising for other Lattice QCD computations.

\section{Performance of a 20M Point 3D U(1)$\sigma$ Model on the 34K Core Prototype}
\label{performance}

To test the speed of the S1 prototype, we ran a moderately large U(1)$\sigma$ model on it, as well as on a traditional CPU, and compared their respective speeds and efficiencies (speed/watt).

We use a similar algorithm to the one explained above. As discussed in section \ref{bcc}, the prototype machine is a 2D SIMD mesh of $192 \times 176 ~(=33,792)$ cores, each of which has 512 16-bit words of memory. On this mesh we were able to implement a $384 \times 352 \times 150 ~(=20,275,200)$ point lattice. This model used all 34K cores, and most of the memory per core, as we will discuss.

Appendix \ref{appendix:S1IntCode} shows the code for the S1 model.  It is a C program with embedded Nova code for the accelerator, as described in section \ref{bcc}.  As we explain the S1 model, it may be helpful to review the Appendix for details and clarifications.

Let us say that the 3 dimensions of the lattice have coordinates called row, column, and height. The rows represent y-coordinates, the columns represent x-coordinates, and the heights represent z-coordinates. Mesh points are therefore indexed by $(r, c, h)$. We will refer to a set of points in the height dimension at a given row and column location as a ``stack''. 

The prototype hardware does not support torus connections at the edges of the mesh. This means the cores, and thus lattice points, at each of the edges cannot directly communicate with their logical neighbor points on the opposite edges. We could use the local communication primitives to propagate information across the mesh, but this would be slow. 

We solve this issue by folding our logical lattice in half both across the vertical midline and the horizontal midline, and interweaving the two halves.  This makes local communication slightly slower, because data must jump across two cores to reach a lattice neighbor, but this is more than compensated for by the full correspondence between locality in the lattice and locality in the mesh, even at the edges.

\begin{figure}
\begin{center}
\includegraphics[scale=0.3]{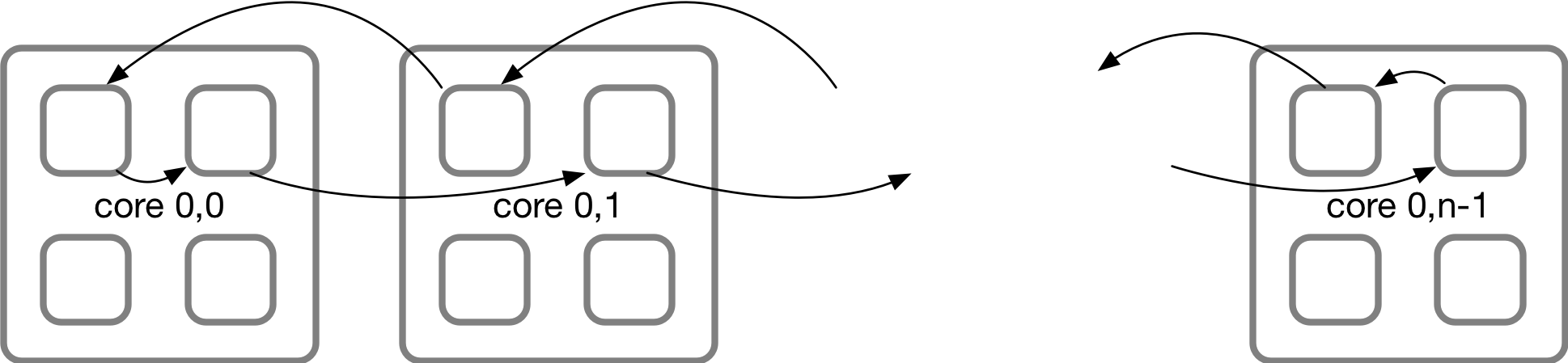}
\end{center}
\caption{Folding and interweaving of points to permit torus communication.} 
\label{fig:folding}
\end{figure}

One approach for folding the lattice is to maintain one lattice point per core (actually, one stack per core) and interweave by placing the left half of the lattice in each row's even cores and the right half of the lattice (in reverse order) in the odd cores, and similarly for the upper and lower halves. However, there is enough memory per S1 core that we chose to place interwoven groups of four lattice points in a single core.

Figure \ref{fig:folding} shows this pattern for a single row of points in a single row of cores.  The arrows show the logical connectivity, that is the nearest neighbors in the lattice, of the lattice points.  The pattern is repeated for both rows of points in each core, for all rows of cores.  An analogous pattern exists in the vertical direction (i.e., columns). Performing a toroidal Get operation requires performing a series of four smaller Get operations, corresponding to the distance 2 and distance 1 links, in each of the westward and eastward (or northward and southward) directions.  For details on these operations, see Appendix \ref{appendix:S1IntCode}.

Each point shown in Figure \ref{fig:folding} actually denotes a stack of lattice points, so each core correspondingly contains four stacks of lattice points. Thus, the 192 rows of cores correspond to a lattice y-dimension of $2 \times 192 = 384$ points, and similarly the x-dimension corresponds to 352 points. 

The mesh height is limited by the 512 words of memory available at each core, at least until we consider using multiple cores per stack as discussed in section \ref{lqcd}.  In order to maximize the height of the mesh, we explored storing spins as single 8 bit integers rather than pairs of 16 bit approximates. We note that in full LQCD models complex numbers are associated with each point, and the representation used here most likely is unsuitable. We discuss full LQCD models further in section \ref{lqcd}.

8 bit integers (i.e., bytes) represent values from 0 to 255. We can represent angles from $0$ to $2\pi$ as integers $i = 256 \times \theta / 2\pi$, where $\theta$ is the angle in radians. 

By representing the angles as single bytes, we can pack two per S1 word, allowing us to double the mesh height. If we choose stacks to be 150 points high, so that each stack is represented by 75 words, all four stacks in a core contain 300 words in total, which is comfortably less than the 512 words available per core.  This leaves space for tables we shall describe, as well as storage needed by the Nova compiler.

An important operation in our algorithm is the dot product. We must take dot products of spins, and since the spins are represented by angles, a natural method of doing so is to use $a \cdot b = |a||b|cos(\theta)$. Since all spins have magnitude 1, this becomes $a \cdot b = cos(\theta)$. 

We thus need to compute cosines, but there is no special hardware support for such functions.  In order to compute cosines efficiently, we pre-compute a table. To minimize the size of the table, we store only cosines for the first quadrant and use the symmetry of the cosine function to find the values in the remaining three quadrants. So the cosine table contains 64 entries.  Further, each cosine entry is a 16-bit integer from -2047 to +2047, representing -1.0 to +1.0.

It also is necessary to find the probability of flipping a spin given the change in energy resulting from a flip. The change in energy may range from -12 to 12, because there are 6 neighbors, and each one could at most flip from +1 to -1 for a change of -2, or the reverse for a change of +2. Since each cosine table entry has magnitude less than 2048, the energy change has maximum magnitude $12 \times 2048 = 24,576$, which fits within a 16-bit signed integer.  If the change in energy is negative, we accept. If it is positive, we accept with a certain probability. We pre-compute and store acceptance probabilities, for changes in energy from 0 to 12, in a 48-entry table indexed by change in energy (rounded to the nearest $1/4$) and then look up the values when necessary. We store the probabilities themselves as integers from 0 to 32,767 and accept if a random number from 0 to 32,767 is less than or equal to the probability. For further detail on the probability representations, see Appendix \ref{appendix:S1IntCode}.

The memory usage of the complete 20M point model is shown in Table \ref{memory}.

\begin{table}
\begin{center}
\begin{tabular}{|p{5.25cm}|c|}
\hline
Purpose & Words Used \\
\hline
\makecell[l]{4 stacks of lattice points, \\ 75 words per stack} & 300 \\
\hline
Cosine table & 64 \\
\hline
Probability table & 48 \\
\hline
Total Used & 412 \\
\hline
Words available for miscellaneous variables and use by Nova compiler & 100 \\
\hline
\end{tabular}
\caption{Memory allocation per core, out of available 512 words, for 150 point high lattice}
\label{memory}
\end{center}
\end{table}

\begin{figure}
\begin{center}
\includegraphics[scale=0.3]{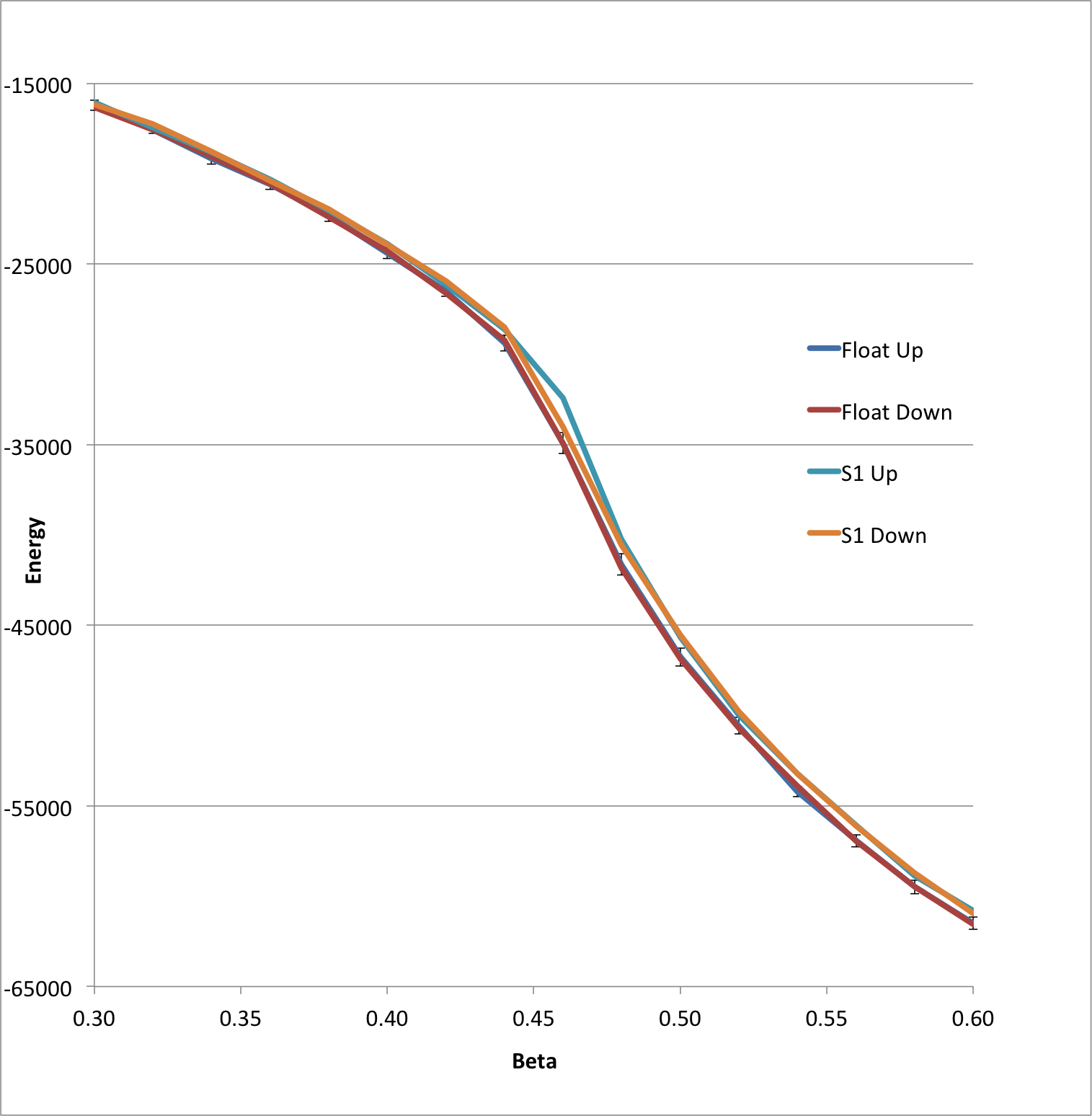}
\end{center}
\caption{Average equilibrium energy per lattice edge (y-axis) vs.~coupling constant $\beta$ (x-axis), for floating point (CPU) and integer (S1) lattice values, over increasing and decreasing coupling constants.  The error bars show one standard deviation above and below the means.}
\label{fig:energyFloatInt}
\end{figure}

We tested the algorithm on a 32 x 32 x 32 point lattice to check its accuracy compared to the floating point model described in section \ref{accuracy}. The results are shown in Figure \ref{fig:energyFloatInt}, and they suggest the MCMC algorithm is correctly implemented on the S1. 

Table \ref{table:performance} shows the speed and efficiency of the full 20M point lattice running on the CPU versus the S1. We show the speed in full mesh updates/second (of either the red or the black points), power usage (measured in watts), updates/second/watt, and relative efficiency between CPUs and the S1. 

In comparing speed and efficiency, the CPU used was a single thread of an Intel Core i7 CPU, implemented using a 22nm silicon process, running at 2.6 GHz. We estimate this processor drew about 20 watts running the model.  The S1 prototype, implemented using a 40nm silicon process, ran at 125 MHz. This S1 system also draws about 20 watts. We calculate efficiency as the ratio updates/second per watt.  First we calculate that ratio using the silicon processes actually used by the respective chips.  Then, to better measure the intrinsic efficiency of each kind of design, we recalculate the ratio as if the S1 were built using the CPU's 22nm silicon process, which gives a more meaningful comparison.  (Note: one would not likely use a 22nm process for a future Singular-style design, because a more modern process, such as 14nm, is probably a better choice.) The S1 design is more efficient that the CPU design by a factor exceeding 2000.

\begin{table}
\begin{center}
\begin{tabular}{|p{4.5cm}|l|l|}
\hline
processor & CPU & S1 \\
\hline
\hline
updates/sec & 0.2 & 153 \\
\hline
relative speedup & & 765$\times$ \\
\hline
\hline
watts & 20 & 20 \\
\hline
updates/sec/watt & 0.01 & 7.65 \\
\hline
relative efficiency & & 765$\times$ \\
\hline
relative efficiency with S1 scaled to 22nm silicon process &  & 2529$\times$ \\
\hline
\end{tabular}
\caption{Performance of S1 vs CPU on $384 \times 352 \times 150$ (= 20 million) point lattice}
\label{table:performance}
\end{center}
\end{table}

\begin{figure}
\begin{center}
\includegraphics[scale=0.16]{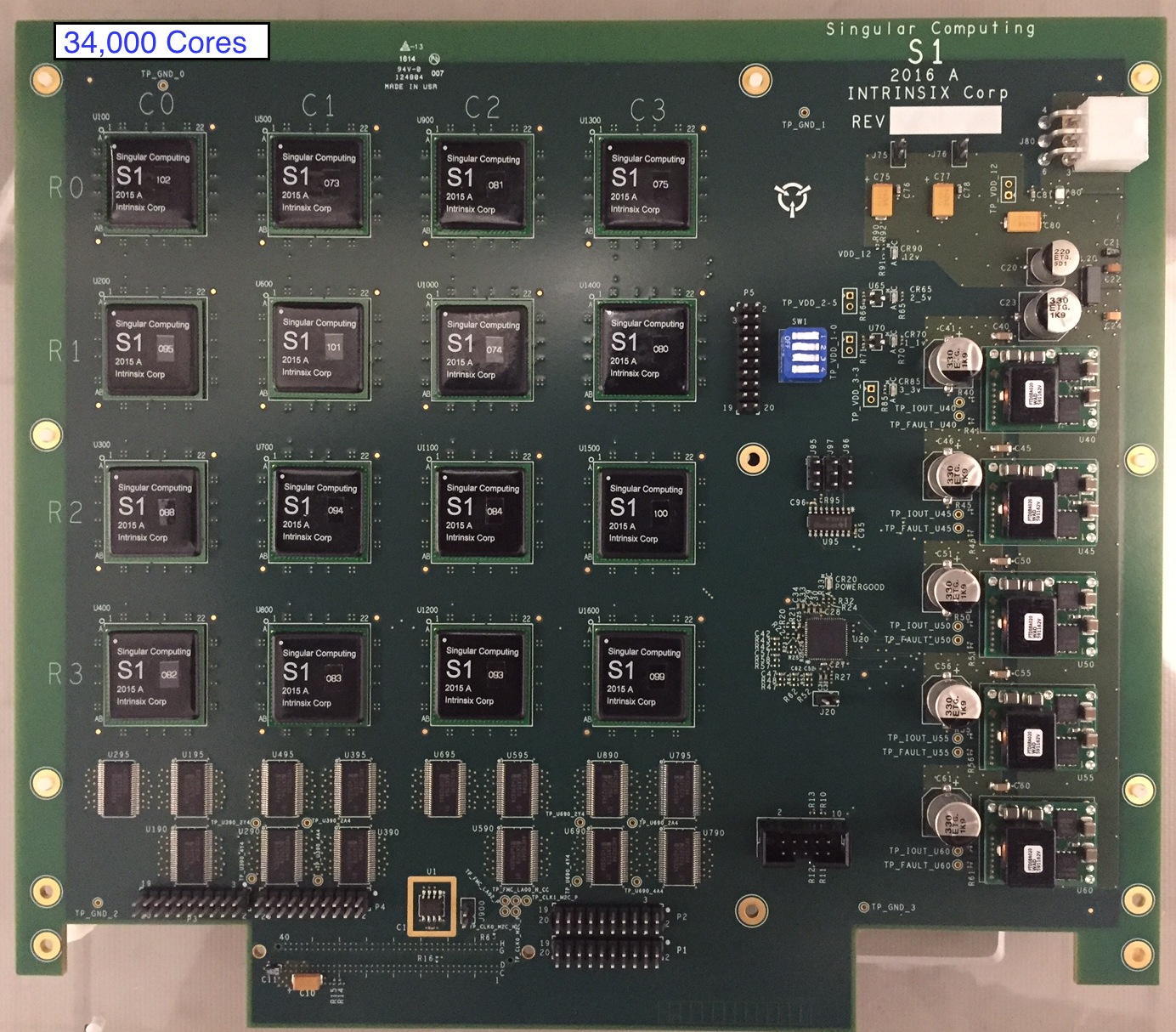}
\end{center}
\caption{34,000 core single-board S1 prototype (12 inch x 10 inch).}
\label{fig:S1Board}
\end{figure}

In conclusion, even using a small, prototype, single-board computer, as shown in Figure \ref{fig:S1Board}, we were able to run a 20 million point U(1)$\sigma$ model. Neither the S1 nor the CPU software was especially optimized, and the S1 prototype hardware was developed at very low cost and is far from optimized, so the results here are merely indicative.  However, these results suggest that approximate mesh machines can operate thousands of times more efficiently than traditional CPUs. This implies that future approximate machines could run U(1)$\sigma$ models quickly over very large lattices, using very little energy, and suggests further research be undertaken to test the speed and accuracy of approximate hardware for full LQCD models.

\section{Extending to LQCD}
\label{lqcd}

We have discussed several implementations of U(1)$\sigma$ models on massively parallel, SIMD, approximate computers, and now wish to offer a few thoughts on extending this work toward full LQCD models.  

It is our understanding that our U(1)$\sigma$ models have several algorithmic patterns in common with LQCD, including red/black checkerboard updates, local communication, parallel random numbers, and conditional execution.

However, there are patterns that seem to be essential to LQCD models (given current understanding and algorithms) that were not used in the U(1)$\sigma$ models, such as the following.

\emph{Matrices instead of spins.}
It is straightforward to store small matrices at each lattice point, instead of simple spins, and operate on them using code similar that in Algorithm \ref{alg:matmul}. The limiting factor is APE memory, which we discuss below along with large high dimensional lattices.

\emph{High precision.}
Our understanding is that the sign problem arises in LQCD, and that in general some parts of traditional LQCD algorithms need high precision arithmetic.  There are various ways to address this issue, using both general and specialized approaches.

The most general approach is that Singular-style machines can perform arbitrary bit manipulations on 16-bit words, as well as doing approximate arithmetic, so we can build high precision integer and floating point libraries operating on multi-word representations. 

However, in order for these machines to operate efficiently on a task, the amount of high precision computation needs to be small compared to the amount of approximate or 16-bit integer computation, so such libraries must be used sparingly.

Singular has found that there are many task-specific ways to improve precision, without using the big hammer of a high precision library (or high precision hardware), and which are much more efficient than using those big hammers.

An example of this is Kahan (or ``compensated'') summation.  This technique was originally invented to improve the quality of long floating point summations, where it can be proven to keep accumulated error low.  For approximate arithmetic it is in practice effective, allowing thousands of values to be summed, despite incurring 1\% error at each operation.

Another example of improving precision at low cost is in high dimensional nearest neighbor calculations.  Given a set of n-vectors and a query vector, we want the vector from the set that has the least distance from the query.  If, for instance, we place each vector from the set in its own core, broadcast the query, ask each core in parallel to calculate its distance, and then do a global minimum reduction to find the nearest vector, we will generally find that if n is large, e.g.\ 1000, then the mesh frequently will return an incorrect (non-nearest) vector, e.g. 20\% of the time (using as an example a set of 10,000 vectors of random values between 0 and 1).  However, if we have the mesh return the two nearest neighbors, according to its approximate calculations, and then have the CPU host determine which of those two vectors is actually nearest, using its high precision (floating point) arithmetic, then the error rate may drop to e.g.\ 5\%.  Reading back the nearest five neighbors, and having the CPU choose among them, can drop the error rate to under 0.1\%.  The conclusion is that we can get good results, with much lower error than the 1\% of the approximate arithmetic, using relatively low overhead in extra CPU computation.

A final example, for solving linear systems, is to use mixed precision iterative refinement algorithms, such as for congugate gradient problems.  Most of the computational work, expressed as inner loops, is performed in low precision approximate arithmetic, while the outer loops use a high precision floating point library running on the APEs. This method can yield, for instance, error $100\times$ lower than 1\% using twice the compute time, and even lower error using more compute time.  Given that the inner loops run hundreds or thousands of times more efficiently than on CPUs, this can be an effective approach.

Singular has found that for many tasks modest creativity leads to solutions like the above, allowing approximate hardware to solve important problems.  Of course the devil is in the details and future experiments seem needed to judge if LQCD can be implemented efficiently and yield high quality results, using task-specific methods, high precision libraries, or some combination.

\emph{Mapping the logical lattice to the physical mesh.}
In the models we described earlier, the logical lattices have similar sizes as the physical hardware mesh.  However, one wants to be able to use lattices that do not exactly match the hardware.

If a lattice dimension is smaller than a hardware dimension, it is easy to use the conditional execution ability of the APEs to turn off APEs that fall beyond the lattice in that dimension.  The total computation of the system then falls below its maximum capacity, but the power shrinks accordingly, so that efficiency remains relatively constant.

If a lattice dimension is larger than a hardware dimension, then one might wish to place multiple adjacent lattice points in a single APE (assuming there is adequate APE memory for this embedding).  Control Unit loops make it easy to have each APE process the multiple lattice points stored within.  Each APE then can be thought of as simulating multiple virtual APEs, thereby expanding the hardware mesh by a multiple in some dimension.

In the above scheme, if the lattice dimension is not an integral multiple of the hardware dimension, then we can use APE conditional execution, in conjunction with Control Unit loops, to have some APEs sleep while other APEs finish their full set of lattice points.

\emph{Large high dimensional lattices.}
Real LQCD models have four or five dimensions.  Machines that expose the physical realities of modern silicon, as discussed in Section \ref{bcc}, essentially are two dimensional.  Therefore, we must compact the higher dimensions into a 2D grid.

For very small lattices, we could imagine mapping each subspace corresponding to a particular $(x,y)$ projection into a single APE located at $(x,y)$.  However, for realistic lattices the subspaces are large and need to be mapped to collections of APEs.

\begin{figure}
\begin{center}
\includegraphics[scale=0.4]{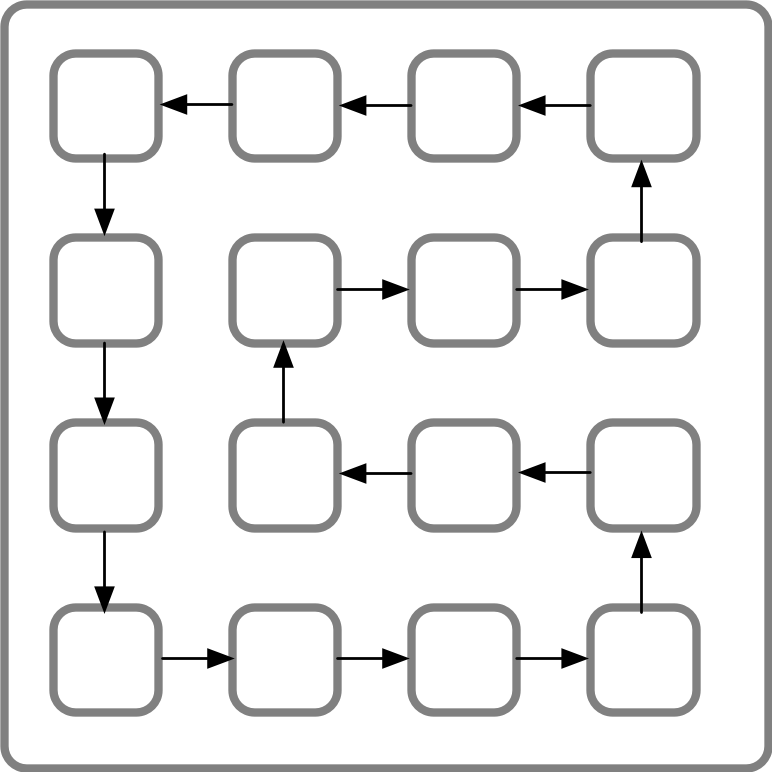}
\end{center}
\caption{``Snake'' pattern of a stack of lattice points mapped into a $4 \times 4$ tile of APEs.}
\label{fig:snake}
\end{figure}

As an example of such a mapping, we built a model in which each stack of a 3D lattice was mapped to a $4 \times 4$ tile of APEs.  Each APE in the tile stored a contiguous sub-sequence of lattice points from the stack.  The APEs in the tile were arranged in a ``snake'' pattern, so that lattice points adjacent in the stack were also adjacent in the snake.  Figure \ref{fig:snake} shows the snake pattern. Appendix \ref{appendix:2dembedded3dising} shows the full implementation.

This mapping means that when a lattice point wants to get the spins of neighbors above or below in its stack, those values are either in the same APE as the point, or are in APEs that are physically adjacent in the snake and thus in the APE mesh.  Points that are adjacent in the row/column directions in the lattice are four APEs away in the row/column directions in the APE mesh.

For 3D lattices of sizes used in LQCD-like models, we anticipate that the delays incurred by mapping stacks into multi-APE tiles, given the data size of the individual lattice points and the memory size of the APEs, is likely not to dominate the computations.  For 4D and 5D lattices, code will need to be written to judge the costs incurred, and confirm that communication time will be reasonable compared to compute time.

\emph{Global and semi-global reductions and broadcasts.}
The S1 prototype performs all communication using local \emph{Get} operations together with global \emph{Read} and broadcast operations.  These capabilities were sufficient for our U(1)$\sigma$ models, but we understand that LQCD codes make significant use of both global and semi-global reductions and scatter/gathers, for example when solving Dirac equations.

While these communication patterns can be implemented using S1 operations, the S2 successor to the S1 is being designed to better support global and semi-global communication.  There are limits to what can be supported, given the essential 2D nature of current silicon chips (or even the limited 3D nature of still speculative technology being developed by semiconductor companies), but there is a reasonable likelihood that the S2 will perform these operations an order of magnitude faster than the S1 for a given clock speed.

Whether this is sufficient to keep communication costs less than compute costs, and thus keep compute efficiency high, needs new code to be developed and remains to be seen.  However, there is a strong sense in which current semiconductor technology can do no better.  The S1/S2, to within a small constant factor, exposes the actual timing of data flow over a 2D chip, and the programmer gets to use silicon's capabilities fully, with little overhead from hardware architecture, to get the job done.

\emph{Machine Learning.} 
The USQCD Collaboration report \emph{Status and Future Perspectives for Lattice Gauge Theory Calculations to the Exascale and Beyond} \cite{cite:exascale} describes machine learning, and in particular deep neural networks, as a technology that has promise to accelerate LQCD computation and enable new research directions. Approximate computing, and the Singular architecture in particular, is well suited to many AI tasks, including deep learning training and inference.  As Singular has shown in public demonstrations, approximate computers can be as efficient as specialized deep learning hardware, such as the Google TPU, while retaining the programability of relatively general purpose devices such as GPUs.

\section{Scaling Up to A Billion Cores}
\label{scaling}

Scaling hardware up to 4 million cores per board is not difficult.  As described, the S1 was made using a 40nm silicon process and $25mm^2$ square chips.  While silicon process ``nodes'', such as 14nm, 7nm, and 5nm, have gradually become marketing terms rather than descriptive of linear scaling factors, there are significant density improvements as these processes evolve.  A 14nm process could comfortably fit 256K cores on a CPU/GPU sized chip.  A $4 \times 4$ grid of such chips, as used in the S1, would yield a 4 million core board drawing hundreds of watts at hundreds of MHz operation.  Such a board would have, in aggregate, several GB of in-core memory with petaop compute, memory access, and inter-core communication speeds.

Building systems from these boards could be easy or difficult, depending on how far one wanted to push the inter-board communication speeds.  Singular's general approach is to provide what is technically reasonable, to minimize the pressure on hardware architects to achieve maximal results by using heroic designs, which consume inordinate resources and unreasonably drive up cost, power, and complexity.  Following this path, it is reasonable to provide billion core systems in several racks, with power and cooling well within limits typical for rack mounted systems.

Developing software for 4M core boards would not be particularly more difficult than developing it for the S1.  Developing it for systems composed of hundreds of such boards would be more difficult, in part because of the need to adapt to inter-board communication delays.  However, young programmers often find it exciting and motivating to be using uniquely powerful computers, such as billion core systems, and we expect they will find it fun to be the first to push software in a new direction.

\section{Conclusion}

We have described four U(1)$\sigma$ models: a $32 \times 32 \times 32$ lattice of floating point XY spins run on a CPU, the same model using approximate valued XY spins, a 20 million point lattice of quantized integer angle spins run on the S1 prototype, and a 3D model that flattens each stack of lattice points across a $4 \times 4$ set of cores in the 2D hardware mesh.

The approximate XY model produced results very similar to the floating point XY model, suggesting that approximate arithmetic might be usable as a hardware primitive for certain aspects of LQCD calculations.  The 20 million point S1 model produced results fairly similar to the floating point XY model, while exhibiting  $\sim\!\!2500\times$ better compute/watt than an Intel CPU when implemented using the same silicon fabrication technology.

The ``3D flattened to 2D'' model indicates how one might implement higher dimensional lattices on a 2D hardware substrate.

The Singular design makes it relatively easy to create individual circuit boards holding millions of cores, drawing a few hundred watts, and providing petaops of flexible compute with petabytes/sec of memory and communication bandwidth.  Several racks composed of such boards would contain a billion cores and provide most of an exaop of computing, with exabyte/sec memory and communication bandwidth, at relatively modest cost and power consumption.

We did not implement the more sophisticated algorithms used in LQCD, but discussed how some of the algorithmic patterns found in those codes (including new patterns, such as deep learning) might be implemented on massively parallel approximate computers.  Determining with confidence whether LQCD could run efficiently and produce high quality results on billion core approximate mesh machines remains unfinished, but based on the results herein, we hope the reader concludes that such a question is worth exploring.

\appendix

\section{Complex matrix multiplication, shifted between cores }
\label{appendix:matmul}
\vspace{-1em}
See arXiv ancillary file \emph{matmul.c}.
\vspace{-.5em}

\section{Floating point and approximate spins model}
\label{appendix:floatApproxCode}
\vspace{-1em}
See arXiv ancillary file \emph{xyModel.c}.
\vspace{-.5em}

\section{S1 integer spin model}
\label{appendix:S1IntCode}
\vspace{-1em}
See arXiv ancillary file \emph{S1xyModel.c}.
\vspace{-.5em}

\section{Model in which each tall Stack of lattice points is mapped to a $4 \times 4$ tile of cores}
\label{appendix:2dembedded3dising}
\vspace{-1em}
See arXiv ancillary file \emph{3dEmbeddedIn2d.c}.

\end{document}